# Energy Efficient Resource Allocation in Federated Fog Computing Networks

Abdullah M. Alqahtani, Barzan Yosuf, Sanaa H. Mohamed,Taisir E.H. El-Gorashi, and Jaafar M.H. Elmirghani, Fellow, IEEE
School of Electronic and Electrical Engineering, University of Leeds, LS2 9JT, United Kingdom

*Abstract*— There is a continuous growth in demand for time sensitive applications which has shifted the cloud paradigm from a centralized computing architecture towards distributed heterogeneous computing platforms where resources located at the edge of the network are used to provide cloud-like services. This paradigm is widely known as fog computing. Virtual machines (VMs) have been widely utilized in both paradigms to enhance the network scalability, improve resource utilization, and energy efficiency. Moreover, Passive Optical Networks (PONs) are a technology suited to handling the enormous volumes of data generated in the access network due to their energy efficiency and large bandwidth. In this paper, we utilize a PON to provide the connectivity between multiple distributed fog units to achieve federated (i.e. cooperative) computing units in the access network to serve intensive demands. We propose a mixed integer linear program (MILP) to optimize the VM placement in the federated fog computing units with the objective of minimizing the total power consumption while considering inter-VM traffic. The results show a significant power saving as a result of the proposed optimization model by up to 52%, in the VM-allocation compared to a baseline approach that allocates the VM requests while neglecting the power consumption and inter-VMs traffic in the optimization framework.

I. INTRODUCTION (HEADING 1)

The demand for time-sensitive applications such as tactile IoT, virtual reality, and connected vehicles is continuously growing. These applications usually generate unprecedented amounts of data that are difficult for centralized cloud architectures to handle and result in high latency, low spare capacity, and increased power consumption. Therefore, researcher are considering a paradigm shift from centralized architectures towards distributed ones. Fog computing was first coined by Cisco to overcome the issue of centralized processing by the cloud and help by considering computing resources closer to end-user devices. However, Fog Computing is characterized by limited computational resources in terms of processing cycles and storage capacity. Therefore, it faces many challenges such as performance, resource allocation and orchestration, and reliability [1]–[8]. Therefore, the capacity of fog units should be addressed to process the intensive end-users demands. The authors in [9] proposed a generic cloud-edge architecture with the aim of facilitating both vertical and horizontal offloading between processing servers. The problem is formulated as a mixed integer nonlinear programming (MINLP) with the objective of optimizing both the workload and capacity during the offloading process.

Moreover, the access network is a critical segment of the network connecting the end-user devices to the Central Office. The end-users as well as enterprise environments constantly require a reliable access network that can provide high bandwidth, low power consumption and low latency. Passive Optical Networks have shown their ability to satisfy such end-user requirements as well as a potential to absorb future growth. PONs have now become a popular network technology in the access and cloud domains due to their energy efficiency and high bandwidths that are particularly suited to time-sensitive application demands [10]–[13]. Typically, the PON architecture includes passive devices such as Arrayed Waveguide grating Routers (AWGR), couplers/splitters, and possibly Fiber Bragg gratings [14].

Vritualization is currently one of the most effective techniques for achieving energy-efficient fog and cloud environments. Virtualization allows multiple virtual machines (VMs) that represent independent applications to run on a single physical machine (server), thus, a Fog computing environment can host a set of distributed applications, each of which runs as a collection of VMs. The VMs are isolated from each other and each VM can utilize a portion of the physical server resources according to the needs of the application by allocating CPU, memory, and bandwidth resources to them [15], [16]. However, some VMs need to communicate with other VMs to complete their tasks. Thus, such VMs could be placed on the same server or on a different server. The traffic between VMs placed in different servers is based on VMs communication that can be considered an "east-west" traffic. The east-west traffic increases the networking power consumption in the data centres environment [17], [18]. Therefore, considering the inter-VMs traffic in the VMs placement optimization is essential in order to achieve an energy-efficient fog computing environment. The authors in [19] investigated the impact of inter-VM traffic on the power consumption . The proposed VM allocation model tackled the cooperation and synchronization communication traffic between VM pairs to optimize the energy efficiency of cloud computing. The authors in [20] show that considering the inter-VMs traffic pattern in the VMs placement model can have a considerable impact on the reduction of the total power consumption.

In this paper, we improve the energy efficiency of distributed federated fog units by optimizing VMs placement whilst taking into consideration the inter-VMs traffic. We benefit from our previous work in energy efficiency that tackled areas such as distributed processing in the IoT/Fog layer [21]–[24], green core and data centre (DC) networks [25]–[34], [35]–[40], network virtualization and service embedding in core and IoT networks [41]–[44] and machine learning and network optimization for healthcare systems [45]–[48] and network coding in the core network [49], [50]. Our previous work in [1] optimized connectivity among distributed fog units to facilitate the communication between different servers. Also, our work in [51] extended the available capacity in fog units by enabling the borrowing of data processing capabilities from nearby fog units to serve intensive VM demands originating from the end user devices. This allowed us to scale up the use of fog processing, while employing a MILP model to optimize the VMs placement taking into consideration the inter VMs traffic. The reminder of this paper is organized as follows: Section 2 explains the proposed architecture and the optimization model. Section 3 presents and discusses the results. Finally, Section 4 concludes the paper and outlines areas for future work.

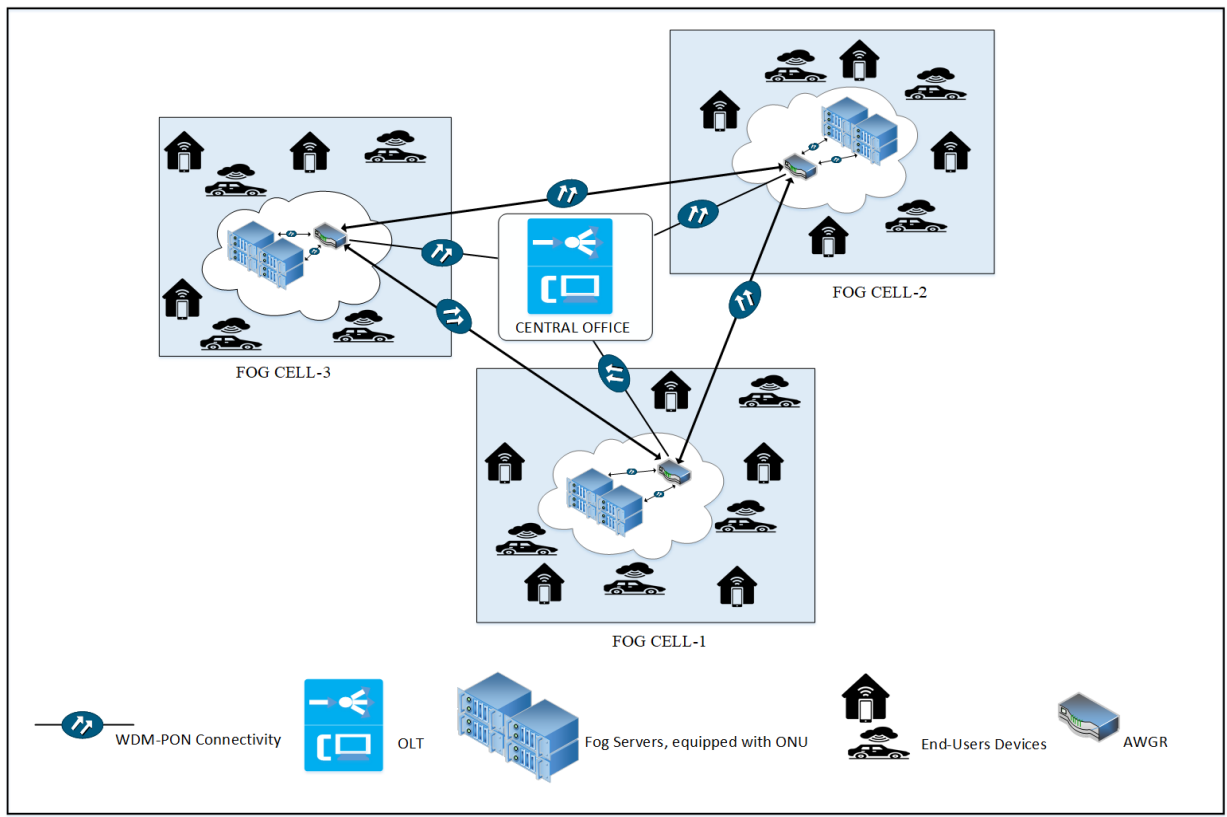

Fig. 1. Federated Fog Computing units over PON

## II. VM PLACMENT OPTIMIZATION FRAMEWORK

### A. The Federated Fog Computing over PON Architecture

As shown in Fig. 1, we consider three distributed fog computing cells connected to each other through WDM-PON in the access network [1]. Each PON cell has two racks of servers, each of which has four servers for hosting the VMs, and each server is equipped with an Optical Network Unit (ONU) acting as a transceiver to send and receive traffic. All three Fog cells share the same Optical Line Terminal (OLT) located in the central office. The OLT is responsible for receiving the VMs requests from the end-user firstly, and is then responsible for placing these VMs based on the resource availability of the fog cells. To make use of the WDM-PON connectivity, each cell has arrayed waveguide grating routers (AWGRs) to achieve full connectivity between racks of servers located within the same fog cell or between racks in different fog cells. The details of the WDM PON connectivity is given in our previous work in [1]. The inter-VM traffic, which flows between VMs placed in different servers, could be traffic flow within the same rack or traffic flow between different racks. Thus, the inter-VM traffic should pass through the AWGR to communicate with other rack(s) within the same cells or in different cells.

### B. MILP Model.

In this section, we develop a MILP model to minimize the total power consumption of the fog architecture shown in Fig. 1, by optimizing the VMs placement subject to processing and networking capacity constraints.

The parameters used in the VM allocation model are as follows:

| S | Set of servers |
|---|---|
| $P_g$ | Set of PON groups (i.e., racks) |
| $P_c$ | Set of PON cells, each one represents a fog computing unit |
| $S_g$ | Set of servers within a PON group. |
| $S_c$ | Set of servers within a PON cell |
| VM | Set of VM requests |
| I | Idle power consumption of the server |
| M | Maximum power consumption of the server |
| C | CPU capacity of the server |
| R | Memory capacity of the server |
| $C_{VM}$ | CPU capacity of VM request |
| $R_{VM}$ | Memory capacity of VM request |
| $T_{if}$ | Traffic demand between VMs $i$ and $f$ { $i$ and $f \in VM$} |

| PO | Power consumption of ONU |
|---|---|
| DO | ONU data rate. |
| W | Wavelength capacity of WDM |
| A | The maximum number of servers that are allowed to serve a VM |
| L | Large number |
| O | Dynamic power consumption range for a server, defined as M - I |

The variables used in the model are as follows:

| $A_s$ | $A_s$ = 1, if server $s$ $\{s \in S\}$ is activated, Otherwise $A_s$ = 0. |
|---|---|
| $W_s^i$ | $W_s^i$ = 1, if request $i$ $\{i \in VM\}$ is processed by server $s$ $\{s \in S\}$, Otherwise $W_s^i$ = 0 |
| $V^{if}$ | $V^{if}$ = 1, if requests $i$ and $f$ $\{i, f \in VM\}$ are placed in the same server $S$ $\{s \in S\}$, Otherwise $V^{if}$ = 0 |
| $T^s$ | The traffic of server $s$ $\{s \in S\}$ |
| $N^s$ | Number of VMs served by server $s$ |
| $Q_{sd}^{if}$ | Defined as the AND of two variables $W_s^i$ and $W_d^f$, $\{i, f \in VM\}$ and $\{s, d \in S\}$. |
| $B_s^{if}$ | Defined as the AND of two variables $\varpi_s^i$ and $W_s^f$, $\{i, f \in VM\}$ and $\{s \in S\}$ |
| $T^{sd}$ | Traffic between two server $s$ and $d$ hosting two $VM$ requests |
| $P_s^i$ | Processing resources used to serve VM, $\{i \in VM\}$, and $\{s \in S\}$. |

The power consumption of resources in fog computing servers is composed of:

1. The power consumption of the activated fog servers ($S_{PC}$) [51]:

$$SPC = \sum_{s \in S} \left( I\ A_s + O \sum_{i \in VM} P_s^i \right) \quad (1)$$

2. The power consumption of ONUs attached to each activated server ($ONU_{PC}$) [51]:

$$ONUPC = \frac{PO}{DO} \sum_{s \in S} T^s \quad (2)$$

The processing resources used to serve VMs are calculated as:

$$P_s^i = \frac{(C_{vm}\ W_s^i)}{C} \quad (3)$$

The MILP model is defined as follows:

**Minimize**:

$$S_{PC} + ONU_{PC} \quad (4)$$

Subject to the following constraints:

$$\sum_{i \in VM} R_{VM}\ W_s^i \leq R, \quad \forall s \in S \quad (5)$$

Constraint (5) ensures that the memory capacity of servers is not exceeded.

$$\sum_{i \in VM} C_{VM}\ W_s^i \leq C, \quad \forall s \in S \quad (6)$$

Constraint (6) ensures that the processing capacity of servers is not exceeded.

$$T^s \leq DO, \quad \forall s \in S \quad (7)$$

Constraint (7) ensures that the total traffic passing through each server does not exceed the data rate of its ONU.

*C. Results and Discussion*

In this section, we evaluate the power savings obtained as a result of the proposed optimized resource allocation model. The model considered three different VM distribution scenarios, which are 10 VMs, 15 VMs, and 20 VMs - with random distributed values for the CPU demand, memory demand and inter-VM traffic data rates as illustrated in Table 3. The VM processing requirements are uniformly distributed across three workloads (10%, 50% and 100%) of the server CPU capacity. The inter-VM traffic demands (i.e. data rates) are distributed uniformly between 100 Mbps and 10 Gbps. Each VM can communicate randomly with up to 4 VMs. The input parameters used in the MILP model are presented in Table 1.

TABLE I. THE INPUT DATA IN MILP MODEL

| Maximum power consumption of the server [52]. | 457W. |
|---|---|
| Idle power consumption of the server. (66% of Maximum power) [52]. | 301.6W. |
| Processing capacity of the server [52]. | 2.5 GHz, 280k MIPS. |
| Processing requirement of the VMs. | (10%, 50% & 100%) of servers' CPU capacity. |
| Memory capacity (RAM) of the server [52]. | 16GB. |
| Memory capacity of the VMs. | 100-500 MB. |
| ONU power consumption [53]. | 2.5W. |
| ONU data rate [53]. | 10 Gbps. |
| Inter-VMs traffic demand. | 100 Mbps-10 Gbps. |
| Capacity of physical link (Wc) . | 60 Gbps |
| Large Number (L). | 1000. |

We have compared the power consumption of the proposed approach for the VM placement problem with a baseline solution. Note that the proposed model's objective function is composed of minimizing the power consumption of the physical machines (servers) that are in charge of hosting the VMs and the power consumption of the ONUs that are in charge of broadcasting and aggregating the traffic flow between servers as a result of inter-VMs traffic. On the other hand, the baseline VM placement model neglects minimizing the processing and networking power consumption, and places the VMs requests without considering inter-VMs traffic demand. Fig. 2 and Fig. 3 present the total power consumption for the three sets of VMs (10 VMs, 15 VMs and 20 VMs) and the number of activated servers needed for hosting these sets. As shown in Fig. 2 the total power consumption is substantially reduced under the proposed approach in comparison to the baseline approach. This is due to the number of activated servers needed to serve the VMs as shown in Fig. 3, the number of servers is fewer with the proposed model. Accordingly, the proposed model with the objective of minimizing the total power consumption has reduced the power consumption of 10 VMs, 15 VMs and 20 VMs by 42%, 52% and 41% respectively in comparison to the baseline. The proposed model places the VMs with high inter-VM traffic in the same server as much as possible in order to reduce the number of activated servers and also to reduce the amount of the traffic traversing to other servers. This results in a reduction of the total power consumption. On the other hand, the baseline VMs placement model places the VMs randomly without considering the inter-VM traffic demand and the workload of the VMs to reduce the total power consumption.

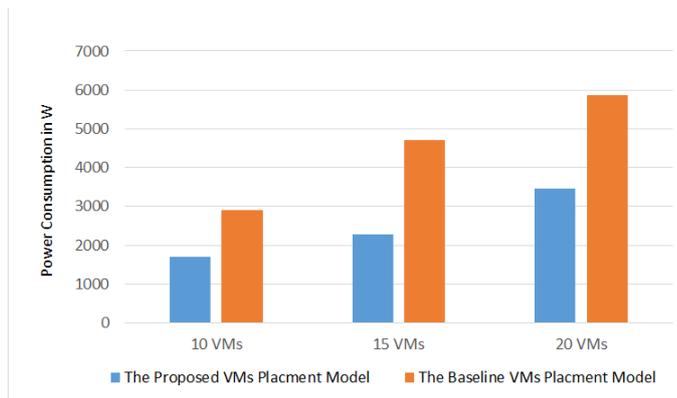

Fig. 2. Power Consumption of proposed model versus baseline VMs placement model

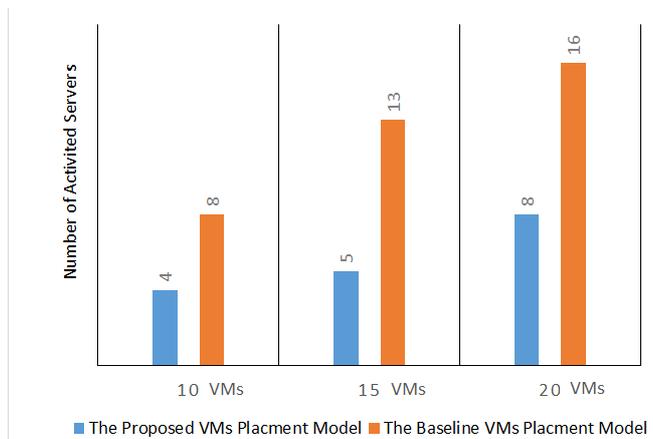

Fig. 3. Number of activated servers in the proposed model versus baseline VMs placement model

III. CONCLUSIONS AND FUTURE WORK

This paper developed a MILP model to optimize the placement of VMs in federated fog computing units over a WDM PON while considering the inter-VM traffic. The results showed power savings of up to 52% compared to a baseline VM-placement model. Future work aims to extend the proposed MILP model to consider a weighted objective function that incorporates delay and power consumption, mobility-aware workload assignment and developing heuristics that are suitable for real-time implementations.

ACKNOWLEDGMENTS

The authors would like to acknowledge funding from the Engineering and Physical Sciences Research Council (EPSRC), INTERNET (EP/H040536/1), STAR (EP/K016873/1) and TOWS (EP/S016570/1) projects. The first author would like to acknowledge the Government of Saudi Arabia and JAZAN University for funding his PhD scholarship. All data are provided in full in the results section of this paper.

REFERENCES

[1] A. M. Alqahtani, S. H. Mohamed, T. E. H. El-Gorashi, and J. M. H. Elmirghani, "PON-Based Connectivity for Fog Computing," in *2020 22nd International Conference on Transparent Optical Networks (ICTON)*, 2020, pp. 1–6.

[2] S. H. Mohamed, T. E. H. El-Gorashi, and J. M. H. Elmirghani, "Energy Efficiency of Server-Centric PON Data Center Architecture for Fog Computing," in *2018 20th International Conference on Transparent Optical Networks (ICTON)*, 2018, pp. 1–4.

[3] A. C. Baktir, A. Ozgovde, and C. Ersoy, "How Can Edge Computing Benefit From Software-Defined Networking: A Survey, Use Cases, and Future Directions," *IEEE Commun. Surv. Tutorials*, vol. 19, no. 4, pp. 2359–2391, 2017.

[4] M. Satyanarayanan, "The Emergence of Edge Computing," *Computer (Long. Beach. Calif).*, vol. 50, no. 1, pp. 30–39, 2017.

[5] K. Dolui and S. K. Datta, "Comparison of edge computing implementations: Fog computing, cloudlet and mobile edge computing," in *2017 Global Internet of Things Summit (GIoTS)*, 2017, pp. 1–6.

[6] S. Yi, Z. Hao, Z. Qin, and Q. Li, "Fog Computing: Platform and Applications," in *2015 Third IEEE Workshop on Hot Topics in Web Systems and Technologies (HotWeb)*, 2015, pp. 73–78.

[7] F. Bonomi, R. Milito, J. Zhu, and S. Addepalli, "Fog computing and its role in the internet of things," *Proc. first Ed. MCC Work. Mob. cloud Comput. - MCC '12*, p. 13, 2012.

[8] Cisco, "Fog Computing and the Internet of Things: Extend the Cloud to Where the Things Are," *WHITE PAPER*, 2015. [Online]. Available: https://www.cisco.com/c/dam/en_us/solutions/trends/iot/docs/computing-overview.pdf. [Accessed: 15-Dec-2019].

[9] M. T. Thai, Y. D. Lin, Y. C. Lai, and H. T. Chien, "Workload and Capacity Optimization for Cloud-Edge Computing Systems with Vertical and Horizontal Offloading," *IEEE Trans. Netw. Serv. Manag.*, vol. 17, no. 1, pp. 227–238, Mar. 2020.

[10] S. H. Mohamed, T. E. H. El-Gorashi, and J. M. H. Elmirghani, "On the energy efficiency of MapReduce shuffling operations in data centers," in *International Conference on Transparent Optical Networks*, 2017.

[11] A. A. Alahmadi, M. O. I. Musa, T. E. H. El-Gorashi, and J. M. H. Elmirghani, "Energy Efficient Resource Allocation in Vehicular Cloud based Architecture," in *21st International Conference on Transparent Optical Networks (ICTON)*, 2019, pp. 1–6.

[12] R. Alani, A. Hammadi, T. E. H. El-Gorashi, and J. M. H. Elmirghani, "PON data centre design with AWGR and server based routing," in *2017 19th International Conference on Transparent Optical Networks (ICTON)*, 2017, pp. 1–4.

[13] J. Baliga, R. Ayre, K. Hinton, and R. S. Tucker, "Energy consumption in wired and wireless access networks," *IEEE Commun. Mag.*, vol. 49, no. 6, pp. 70–77, 2011.

[14] R. Ramaswami, K. Sivarajan, and G. Sasaki, *OPTICAL NETWORK A Practical Perspective*. 2009.

[15] Y. Jin, Y. Wen, and Q. Chen, "Energy efficiency and server virtualization in data centers: An empirical investigation," in *Proceedings - IEEE INFOCOM*, 2012, pp. 133–138.

[16] A. Varasteh and M. Goudarzi, "Server Consolidation Techniques in Virtualized Data Centers: A Survey," *IEEE Syst. J.*, vol. 11, no. 2, pp. 772–783, Jun. 2017.

[17] S. Das, F. Slyne, A. Kaszubowska, and M. Ruffini, "Virtualized EAST-WEST PON architecture supporting low-latency

communication for mobile functional split based on multiaccess edge computing," *J. Opt. Commun. Netw.*, vol. 12, no. 10, pp. D109–D119, Oct. 2020.

[18] S. N. Mthunzi, E. Benkhelifa, M. A. Alsmirat, and Y. Jararweh, "Analysis of VM communication for VM-based cloud security systems," in *2018 Fifth International Conference on Software Defined Systems (SDS)*, 2018, pp. 182–188.

[19] H. A. Alharbi, T. E. H. Elgorashi, A. Q. Lawey, and J. M. H. Elmirghani, "The Impact of Inter-Virtual Machine Traffic on Energy Efficient Virtual Machines Placement," in *2019 IEEE Sustainability through ICT Summit, StICT 2019*, 2019.

[20] X. Meng, V. Pappas, and L. Zhang, "Improving the scalability of data center networks with traffic-aware virtual machine placement," *Proc. - IEEE INFOCOM*, 2010.

[21] B. A. Yosuf, M. Musa, T. Elgorashi, and J. Elmirghani, "Energy Efficient Distributed Processing for IoT," *arXiv Prepr. arXiv2001.02974*, 2020.

[22] S. H. Mohamed, M. B. A. Halim, T. E. H. Elgorashi, and J. M. H. Elmirghani, "Fog-Assisted Caching Employing Solar Renewable Energy and Energy Storage Devices for Video on Demand Services," *IEEE Access*, vol. 8, pp. 115754–115766, 2020.

[23] Z. T. Al-Azez, A. Q. Lawey, T. E. H. El-Gorashi, and J. M. H. Elmirghani, "Energy Efficient IoT Virtualization Framework With Peer to Peer Networking and Processing," *IEEE Access*, vol. 7, pp. 50697–50709, 2019.

[24] H. A. Alharbi, T. E. H. Elgorashi, and J. M. H. Elmirghani, "Energy efficient virtual machines placement over cloud-fog network architecture," *IEEE Access*, vol. 8, pp. 94697–94718, 2020.

[25] A. M. Al-Salim, A. Q. Lawey, T. E. H. El-Gorashi, and J. M. H. Elmirghani, "Energy Efficient Big Data Networks: Impact of Volume and Variety," *IEEE Trans. Netw. Serv. Manag.*, vol. 15, no. 1, pp. 458–474, Mar. 2018.

[26] A. Al-Salim, T. El-Gorashi, A. Lawey, and J. Elmirghani, "Greening Big Data Networks: Velocity Impact," *IET Optoelectron.*, vol. 12, Nov. 2017.

[27] X. Dong, T. El-Gorashi, and J. M. H. Elmirghani, "Green IP over WDM networks with data centers," *J. Light. Technol.*, vol. 29, no. 12, pp. 1861–1880, 2011.

[28] H. M. M. Ali, T. E. H. El-Gorashi, A. Q. Lawey, and J. M. H. Elmirghani, "Future Energy Efficient Data Centers with Disaggregated Servers," *J. Light. Technol.*, vol. 35, no. 24, pp. 5361–5380, Dec. 2017.

[29] N. I. Osman, T. El-Gorashi, L. Krug, and J. M. H. Elmirghani, "Energy-efficient future high-definition TV," *J. Light. Technol.*, vol. 32, no. 13, pp. 2364–2381, Jul. 2014.

[30] A. Q. Lawey, T. E. H. El-Gorashi, and J. M. H. Elmirghani, "BitTorrent content distribution in optical networks," *J. Light. Technol.*, vol. 32, no. 21, pp. 3607–3623, Nov. 2014.

[31] A. Q. Lawey, T. E. H. El-Gorashi, and J. M. H. Elmirghani, "Distributed energy efficient clouds over core networks," *J. Light. Technol.*, vol. 32, no. 7, pp. 1261–1281, Apr. 2014.

[32] J. M. H. Elmirghani *et al.*, "GreenTouch GreenMeter Core Network Energy Efficiency Improvement Measures and Optimization [Invited]," *IEEE/OSA J. Opt. Commun. Netw.*, vol. 10, no. 2, 2018.

[33] M. O. I. Musa, T. E. H. El-Gorashi, and J. M. H. Elmirghani, "Bounds on GreenTouch GreenMeter Network Energy Efficiency," *J. Light. Technol.*, vol. 36, no. 23, pp. 5395–5405, Dec. 2018.

[34] X. Dong, T. E. H. El-Gorashi, and J. M. H. Elmirghani, "On the energy efficiency of physical topology design for IP over WDM networks," *J. Light. Technol.*, vol. 30, no. 12, pp. 1931–1942, 2012.

[35] B. G. Bathula, B. G. Bathula, M. Alresheedi, and J. M. H. Elmirghani, "Energy Efficient Architectures for Optical Networks."

[36] B. G. Bathula and J. M. H. Elmirghani, "Energy efficient Optical Burst Switched (OBS) networks," in *2009 IEEE Globecom Workshops, Gc Workshops 2009*, 2009.

[37] T. E. H. El-Gorashi, X. Dong, and J. M. H. Elmirghani, "Green optical orthogonal frequency-division multiplexing networks," *IET Optoelectron.*, vol. 8, no. 3, pp. 137–148, 2014.

[38] X. Dong, T. El-Gorashi, and J. M. H. Elmirghani, "IP over WDM networks employing renewable energy sources," *J. Light. Technol.*, vol. 29, no. 1, pp. 3–14, 2011.

[39] X. Dong, A. Lawey, T. El-Gorashi, and J. Elmirghani, "Energy-efficient core networks," *2012 16th Int. Conf. Opt. Netw. Des. Model. ONDM 2012*, Apr. 2012.

[40] H. A. Alharbi, T. E. H. Elgorashi, and J. M. H. Elmirghani, "Impact of the Net Neutrality Repeal on Communication Networks," *IEEE Access*, pp. 1–1, Mar. 2020.

[41] L. Nonde, T. E. H. El-Gorashi, and J. M. H. Elmirghani, "Energy Efficient Virtual Network Embedding for Cloud Networks," *J. Light. Technol.*, vol. 33, no. 9, pp. 1828–1849, May 2015.

[42] H. Q. Al-Shammari, A. Q. Lawey, T. E. H. El-Gorashi, and J. M. H. Elmirghani, "Service Embedding in IoT Networks," *IEEE Access*, vol. 8, pp. 2948–2962, 2020.

[43] A. N. Al-Quzweeni, A. Q. Lawey, T. E. H. Elgorashi, and J. M. H. Elmirghani, "Optimized Energy Aware 5G Network Function Virtualization," *IEEE Access*, vol. 7, pp. 44939–44958, 2019.

[44] H. Q. Al-Shammari, A. Q. Lawey, T. E. H. El-Gorashi, and J. M. H. Elmirghani, "Resilient Service Embedding in IoT Networks," *IEEE Access*, vol. 8, pp. 123571–123584, 2020.

[45] M. S. Hadi, A. Q. Lawey, T. E. H. El-Gorashi, and J. M. H. Elmirghani, "Patient-Centric HetNets Powered by Machine Learning and Big Data Analytics for 6G Networks," *IEEE Access*, vol. 8, pp. 85639–85655, 2020.

[46] M. Hadi, A. Lawey, T. El-Gorashi, and J. Elmirghani, "Using Machine Learning and Big Data Analytics to Prioritize Outpatients in HetNets," in *INFOCOM 2019 - IEEE Conference on Computer Communications Workshops, INFOCOM WKSHPS 2019*, 2019, pp. 726–731.

[47] M. S. Hadi, A. Q. Lawey, T. E. H. El-Gorashi, and J. M. H. Elmirghani, "Patient-Centric Cellular Networks Optimization Using Big Data Analytics," *IEEE Access*, vol. 7, pp. 49279–49296, 2019.

[48] I. S. B. M. Isa, T. E. H. El-Gorashi, M. O. I. Musa, and J. M. H. Elmirghani, "Energy Efficient Fog-Based Healthcare Monitoring Infrastructure," *IEEE Access*, vol. 8, pp. 197828–197852, 2020.

[49] M. Musa, T. Elgorashi, and J. Elmirghani, "Bounds for energy-efficient survivable IP over WDM networks with network coding," *J. Opt. Commun. Netw.*, vol. 10, no. 5, pp. 471–481, May 2018.

[50] M. Musa, T. Elgorashi, and J. Elmirghani, "Energy efficient survivable IP-Over-WDM networks with network coding," *J. Opt. Commun. Netw.*, vol. 9, no. 3, pp. 207–217, Mar. 2017.

[51] A. M. Alqahtani, B. Yosuf, S. H. Mohamed, T. E. H. El-Gorashi, and J. M. H. Elmirghani, "Energy Minimized Federated Fog Computing over Passive Optical Networks," May 2021.

[52] Dell, "SPECpower_ssj2008." [Online]. Available: https://www.spec.org/power_ssj2008/results/res2017q3/power_ssj2008-20170829-00780.html. [Accessed: 30-Jan-2020].

[53] K. Grobe, M. Roppelt, A. Autenrieth, J. P. Elbers, and M. Eiselt, "Cost and energy consumption analysis of advanced WDM-PONs," *IEEE Commun. Mag.*, vol. 49, no. 2, pp. 25–32, 2011.